\documentclass[11pt,a4paper]{article}
\pdfoutput=1
\usepackage{jheppubnohead}
\usepackage[utf8]{inputenc}
\usepackage{amsmath}
\usepackage{epsfig}
\usepackage{graphicx}
\usepackage{amssymb}
\usepackage{tabularx}
\usepackage[normalem]{ulem} 
\usepackage{booktabs} 
\usepackage{subfig}
\providecommand{\proarrow}[0]{\rightarrow}

\providecommand{\dif}[0]{\mathrm{d}}

\providecommand{\proname}[2]{#1 \proarrow #2}

\providecommand{\abs}[1]{\left\lvert #1 \right\rvert}

\providecommand{\abss}[1]{\left\lvert #1 \right\rvert^2}

\providecommand{\mire}[1]{{\rm Re} \left[ #1 \right]}

\providecommand{\order}[1]{{\cal O} \left( #1 \right)}
\providecommand{\torder}[1]{{\cal O} \bigl( #1 \bigr)}

\providecommand{\traza}[1]{{\rm Tr} \left( #1 \right)}

\providecommand{\cpm}[0]{\mathcal{M}}

\providecommand{\mmed}[0]{M}

\providecommand{\emed}[0]{E}

\providecommand{\rf}[0]{r}
\providecommand{\pra}[0]{\tilde{M}_1}
\providecommand{\prb}[0]{\tilde{M}_2}
\providecommand{\pia}[0]{\tilde{\Gamma}_1}
\providecommand{\pib}[0]{\tilde{\Gamma}_2}
\providecommand{\conmut}[2]{[#1,#2]}
\providecommand{\anticonmut}[2]{\{#1,#2\}}
\providecommand{\rhop}[0]{\rho_p}
\newcommand{\be}{\begin{equation}}
\newcommand{\ee}{\end{equation}}
\newcommand{\bea}{\begin{eqnarray}}
\newcommand{\eea}{\end{eqnarray}}

\title{On different approaches to freeze-in and freeze-out leptogenesis with quasi-degenerate neutrinos}

\author[]{J.~Racker}

\affiliation[]{Instituto de Astronom\'{\i}a Te\'orica y Experimental (IATE),  Universidad Nacional
de C\'ordoba (UNC)~- Consejo Nacional de Investigaciones Cient\'{\i}ficas y T\'ecnicas
(CONICET) \\ Laprida 854, X5000BGR,  C\'ordoba, Argentina.\\
Observatorio Astron\'omico de C\'ordoba (OAC), Universidad Nacional de C\'ordoba (UNC) \\ Laprida 854, X5000BGR, C\'ordoba, Argentina.
}

\abstract{We compare two approaches for determining the generation of lepton asymmetry during production and decay of quasi-degenerate neutrinos, namely the density matrix formalism and a recent proposal which does not involve any counting of neutrino number densities and is based on plugging the resummed propagator in a quantum field theory model for neutrino oscillations. We show numerically and analytically that they are almost equivalent except in the limit of very large mass splittings. The comparison, performed in a simple scalar toy model, helps to understand several issues that have been discussed in the literature.}
\begin{document}

\maketitle
%%%%%%%%%%%%%%%%%%%%%%%%%%%%%%%%%%%%%%%%%%%%%%%%%%%%%%%%%%%%%%%%%%%%%
%%%%%%%%%%%%%%%%%%%%%%%%%%%%%%%%%%%%%%%%%%%%%%%%%%%%%%%%%%%%%%%%%%%%%%
\section{Introduction}

Quasi-degenerate exotic neutrinos allow for low-scale leptogenesis mechanisms that may be tested in current and near future experiments. More specifically, in the type I seesaw model the baryon asymmetry of the universe can be generated in the freeze-out of Majorana neutrinos with typical masses around $\order{10^2-10^3~{\rm GeV}}$, and also in the freeze-in of much lighter neutrinos, with masses even below the GeV scale. The former mechanism is commonly known as resonant leptogenesis~\cite{pilaftsis03} and the latter as baryogenesis via neutrino oscillations (or ARS leptogenesis)~\cite{akhmedov98, asaka05}. Both mechanisms have been analyzed with different formalisms, see e.g.~\cite{Dev:2017wwc},~\cite{Drewes:2017zyw}, and section II of~\cite{Klaric:2021cpi} for comprehensive reviews and references. 

Indeed the treatment of CP violation in baryogenesis models with quasi-degenerate neutrinos and its implementation in transport equations is not trivial and has actually been discussed over several decades. Some points of concern include the proper calculation of the CP asymmetry in neutrino decays to avoid a divergence in the limit of equal masses (and particularly in the double degenerate limit of equal masses and couplings), the distinction and interplay of two sources of CP violation (mixing and oscillations), and the joint analysis of freeze-in and freeze-out leptogenesis with a single set of kinetic equations valid for any spectrum of the heavy neutrino masses (cf.~\cite{Fidler:2011yq, garbrecht11, garny11, Garbrecht:2014aga, Dev:2014laa, Dev:2015wpa, Dev:2014wsa, Kartavtsev15, Klaric:2020lov, Racker20, Klaric:2021cpi, Jukkala:2021cys, Racker21, Hernandez:2022ivz}). As a matter of fact, there are differences in the literature regarding these issues, like in the regulators for the decay asymmetry (reviewed e.g. in~\cite{Dev:2017wwc}),  in the relative sign of the contributions from mixing and oscillations (compare~\cite{Dev:2014laa, Dev:2015wpa, Dev:2014wsa, Kartavtsev15} with~\cite{Racker20, Jukkala:2021cys, Racker21}), in the existence of an interference term between mixing and oscillations (compare~\cite{Dev:2014laa, Dev:2015wpa, Dev:2014wsa} with~\cite{Kartavtsev15, Racker20, Racker21}), and in the set of kinetic equations (cf.~\cite{Dev:2014laa, Klaric:2020lov}). Note in particular that for a hierarchical spectrum of neutrino masses, the standard classical approach requires the inclusion of real intermediate state (RIS) subtracted rates to be consistent with unitarity, which are present in~\cite{Dev:2014laa} but not in~\cite{Klaric:2020lov} (although the correct hierarchical limit is claimed to be obtained in both works, see also~\cite{Racker20, Jukkala:2021cys, Racker21}). 

Motivated by these considerations we compare in this work two different approaches to freeze-in and freeze-out leptogenesis, namely the one based on quantum kinetic equations for matrices of densities and the one proposed in~\cite{Racker20}. 
The density matrix approach (DMA), like the one used in~\cite{Klaric:2020lov, Klaric:2021cpi} to study jointly freeze-in and freeze-out leptogenesis, is based on a generalization of the Sigl and Raffelt formalism~\cite{Sigl93} to include heavy neutrinos, and can also be derived under some approximations from non-equilibrium quantum field theory (see e.g. ~\cite{garbrecht11, Garbrecht:2014aga, Drewes:2016gmt, Jukkala:2021cys}).  In the approach proposed in~\cite{Racker20}, an expansion around the complex poles of the resummed one-loop propagator is plugged into a quantum field theory model of neutrino oscillations, in order to obtain time dependent probabilities for lepton number violating processes. The CP asymmetry obtained from these probabilities is suitably integrated over time in order to get a source term for the evolution of the lepton asymmetry. Given that in~\cite{Racker20, Racker21} we have employed an external wave packet model for neutrino oscillations~\cite{1963AnPhy22, PhysRevD.48.4310} (following~\cite{Beuthe01}), we will refer to it as the external wave packet approach (EWPA). The EWPA is particularly transparent regarding unitarity (which is a key principle to guide understanding), because no counting of the heavy--unstable--neutrino states is performed.

The issues we are interested in discussing can be captured in a simple scalar toy model, consequently we organize this work as follows: In section~\ref{section_model} we review the main results from~\cite{Racker20, Racker21} and work out the kinetic equations in the DMA for the scalar toy model, then in section~\ref{section_comparison} we compare both approaches and discuss the results, finally we conclude in section~\ref{section_conclusions}.

%%%%%%%%%%%%%%%%%%%%%%%%%%%%%%%%%%%%%%%%%%%%%%%%%%%%%%%%%%%%%%%%%%%
%%%%%%%%%%%%%%%%%%%%%%%%%%%%%%%%%%%%%%%%%%%%%%%%%%%%%%%%%%%%%%%%%%%%%
\section{Implementation of two approaches in a scalar toy model}
\label{section_model}
We will consider a simple scalar toy model which has often been used to study basic aspects of leptogenesis with quasi-degenerate neutrinos.
The particle content is given by one complex and two real scalar fields, denoted by $b$ and $\psi_i$ ($i=1,2$), respectively. The Lagrangian, in the basis where the mass matrix of the real scalars is diagonal, can be written as
\begin{equation}
\label{equation_lag}
\mathcal{L} = \frac{1}{2} \partial^\mu \psi_i \, \partial_\mu \psi_i - \frac{1}{2} \psi_i M_i^2 \psi_i +  \partial^\mu \bar b \, \partial_\mu b - m^2 \, \bar b b - \frac{h_i}{2} \psi_i \, b b - \frac{h_i^*}{2} \psi_i \, \bar b \bar b - \frac{\lambda}{2\cdot 2}(\bar b b)^2 \; .
\end{equation}

The $b$-particles play in this toy model the analogous role that leptons play in standard leptogenesis, hence they will be referred to as ``leptons'', whose mass $m$ will be neglected for simplicity. Lepton number is violated by the cubic Yukawa interaction terms involving the $\psi_i$, to be called ``neutrinos'' henceforth. The last term is a quartic interaction which does not violate lepton number, therefore it will not appear explicitly in our analysis, although we note that it could be used as a way to localize the leptons and thus satisfy the conditions to have oscillations~\cite{Beuthe01}.

Next we describe two approaches that can be implemented to study the production of lepton symmetry in this model.

\subsection{External wave packet approach}
The EWPA, proposed in~\cite{Racker20} and extended to highly degenerate states in~\cite{Racker21}, proceeds in three steps:  (1) expansion around the complex poles of the resummed propagator, (2) calculation of a time dependent CP asymmetry between lepton number violating processes using a quantum field theory model for neutrino oscillations, and (3) integration over time of this CP asymmetry to obtain a source term for the evolution of the lepton asymmetry. Below we extract the main formulae from~\cite{Racker21} that will be needed in this work.

The resummed one-loop propagator matrix, ${\bf G}$, is given by $i {\bf G}^{-1}(p^2) = p^2 {\bf 1} - {\bf M^2}(p^2)$, with 
 \begin{equation}
 \label{ecua_:msquare}
 {\bf M^2}(p^2)= 
\begin{pmatrix}
M_1^2 + \Sigma_{11}(p^2)  & \Sigma_{12}(p^2) \\
\Sigma_{21}(p^2)  & M_2^2 + \Sigma_{22}(p^2) 
\end{pmatrix},
\end{equation}
and
\begin{eqnarray}
\label{ecua_:selfe}
\Sigma_{ii}(p^2)&=&\frac{\abss{h_i}}{(4\pi)^2} \left[1 + \ln \frac{p^2}{M_i^2} - \frac{p^2}{M_i^2} - i \pi \right]  \quad ({\rm for} \; i=1,2) \, ,\\
\label{equation_sigma12}
\Sigma_{12}(p^2)&=&\Sigma_{21}(p^2)=\frac{\mire{h_1^* h_2}}{(4\pi)^2} \left[\frac{M_2^2 \ln \frac{p^2}{M_1^2} - M_1^2 \ln \frac{p^2}{M_2^2} - p^2 \ln \frac{M_2^2}{M_1^2}}{M_2^2-M_1^2} - i \pi \right] .
\end{eqnarray}
The propagator can be expanded around the two complex poles following~\cite{fuchs16}, namely  ${\bf G} \simeq {\bf Z}^{T} \, {\bf \Delta}^{\rm BW} \, {\bf Z}$, with
\begin{equation}
\label{ecua_:breitwig}
{\bf \Delta}^{\rm BW} = i
\begin{pmatrix}
(p^2 - \cpm_{a}^2)^{-1} & 0 \\
0 & (p^2 - \cpm_{b}^2)^{-1} 
\end{pmatrix} , \quad 
{\bf Z} = 
\begin{pmatrix}
\sqrt{Z_1} &  \sqrt{Z_1} Z_{12}\\
\sqrt{Z_2} Z_{21} & \sqrt{Z_2}
\end{pmatrix}, 
\end{equation}
and $\cpm_{a,b}^2$ the poles of the propagator given by the roots of the determinant of ${\bf G}^{-1}$.

At $\torder{h^2}$ (with $h$ representing any of the Yukawa couplings), the elements of the ${\bf Z}$ matrix are equal to
\begin{eqnarray*}
Z_{12} &=& -\theta' + \order{h^4} = - Z_{21} + \order{h^4} \, ,\notag \\ 
Z_1 &=& \frac{1+\rf}{2 \rf}  + \order{h^4} = Z_2  + \order{h^4}\, , 
\end{eqnarray*}
with
\begin{equation*}
\rf \equiv \sqrt{1-\left[ \frac{2 \mire{h_1^* h_2}}{16 \pi \epsilon + i \left(\abss{h_1} - \abss{h_2} \right)} \right]^2} \; , - \theta' \equiv  \frac{i \, \mire{h_1^* h_2}/(16 \pi)}{\epsilon+ i \left(M_1 \Gamma_1 - M_2 \Gamma_2 \right)} \, \frac{2}{1+\rf} \, ,
\end{equation*}
$\epsilon \equiv M_2^2 - M_1^2$, and the $\order{h^4}$ terms are finite in the limit $\epsilon \to 0$. Taking
\begin{eqnarray*}
\cpm_a^2 &=& M_1^2 - i M_1 \Gamma_1 + i \, \frac{\mire{h_1^* h_2}}{16 \pi} \, \theta' \, \equiv \pra^2 - i  \pra \pia \, , \notag\\
\cpm_b^2 &=& M_2^2 - i M_2 \Gamma_2 - i \, \frac{\mire{h_1^* h_2}}{16 \pi} \, \theta' \, \equiv \prb^2 - i  \prb \pib \, ,
\end{eqnarray*}
with $\Gamma_i = \tfrac{\abss{h_i}}{16 \pi M_i} \; (i=1,2)$, it can be verified that $|{\bf G}^{-1} (\cpm_{a,b}^2)| = 0 + \order{h^8, h^6 \epsilon}$. The last equalities in each of the above equations define the real quantities $\pra, \prb, \pia$ and $\pib$.

Using the approximate propagator in an external wave packet model of neutrino oscillations and neglecting the factors related to coherence and localization (which could destroy oscillations), one arrives at these time-dependent probabilities for the lepton number violating processes (see~\cite{Racker21}):
\begin{eqnarray}
\label{ecua_:prob}
\abss{A}(t)  &=&  N \left|\left( h_1^{* 2} - 2 h_1^* h_2^* \,\theta' + h_2^{* 2} \,\theta'^{\,2}\right)  e^{-i \left(\pra - i \frac{\pia}{2}\right) \frac{t}{\gamma}} \; + \notag \right.\\ & & \quad \; \left. \left( h_2^{* 2} + 2 h_1^* h_2^* \,\theta' + h_1^{* 2} \,\theta'^{\,2} \right)    e^{-i \left(\prb - i \frac{\pib}{2}\right) \frac{t}{\gamma}}  \right|^2 \, , \notag \\
\abss{\bar A}(t)  &=&   N \left|  \left( h_1^2 - 2 h_1 h_2 \,\theta' + h_2^2\, \theta'^{\,2} \right) e^{-i \left(\pra - i \frac{\pia}{2}\right) \frac{t}{\gamma}} \; + \notag \right.\\ & & \quad \; \left.   \left( h_2^2 + 2 h_1 h_2 \,\theta' + h_1^2\, \theta'^{\,2} \right) e^{-i \left(\prb - i \frac{\pib}{2}\right) \frac{t}{\gamma}}  \right|^2 \, ,
\end{eqnarray}
where $A \equiv A(\proname{\bar b \bar b}{b b})$, $\bar A \equiv A(\proname{b b}{\bar b \bar b})$, $N$ is a normalization constant, $t$ is the time between production and decay of the neutrinos that mediate these processes, $\gamma \equiv \emed/\mmed$ is the Lorentz factor, $\emed$ is the average energy of the neutrinos, and $\mmed \equiv (M_1 + M_2)/2$.

Finally, a proper integration over time of the  CP asymmetry $\abss{A} - \abss{\bar A}$ yields a source term, $S_{\rm EWPA}$, for the evolution of the lepton density asymmetry $n_L \equiv n_b - n_{\bar b}$, where $n_b$ ($n_{\bar b}$) is the number density of leptons (antileptons). For our purposes it is sufficient to consider a static universe, a single average momentum $p$ for the neutrinos (to avoid unessential integrals over momentum), and neglect finite density effects. Hence the normalization constant $N$ in eq.~\eqref{ecua_:prob} is the same as in~\cite{Racker21}~\footnote{The factor $\abss{Z_1}$ was omitted by mistake in some of the equations of~\cite{Racker21}.}, $N=\abss{Z_1}/(32 \pi \emed)^2$ and the evolution of $n_L$ is given by 
\begin{equation}
 \label{ecua_:sewpa}
 \frac{\dif n_L}{\dif t} = S_{\rm EWPA}(t) - W(t) =  2 \int_0^t n^{\rm eq}(t') \, \Big(\abss{A(t-t')} - \abss{\bar A(t-t')} \Big) \, \dif t' \; - W(t) \, .
 \end{equation}
 The source, $S_{\rm EWPA}(t)$, contains the terms in the kinetic equation which may be non-null in the absence of a lepton density asymmetry and will be the focus of this work, while the washout part, $W(t)$, will not be considered. Moreover, $n^{\rm eq}(t)$ is the equilibrium number density of a scalar particle with mass $\mmed$. Note that in an expanding universe, $n^{\rm eq}(t)$ varies according to the time-dependent temperature, while for our simplified study in a static universe $n^{\rm eq}$ will be varied artificially, with equilibrium corresponding to constancy over time.

\subsection{Density matrix approach}
Following~\cite{Sigl93} we can obtain a kinetic equation for the density matrix $\rho$ of the neutrinos and write it as
\begin{equation}
\label{ecua_:rho}
\frac{\dif \rho}{\dif t} = -i \conmut{H}{\rho} -\frac{1}{2} \anticonmut{\Gamma^d}{\rho} + \gamma^p \, ,
\end{equation}  
where each term depends on the momentum $k$ of the neutrinos, $\Gamma^d$ is a decay rate matrix, $\gamma^p$ is the production term and we will work in the basis where the Hamiltonian is diagonal, $H=\rm{Diag}(E_1,E_2)$, with $E_i=\sqrt{k^2+M_i^2}$ ($i=1,2$). Note that for simplicity we have used Maxwell-Boltzmann statistics and we are not including terms proportional to the lepton asymmetry. In the scalar model the decay rate matrix is given by
\begin{equation}
\label{ecua_:rated}
\Gamma^d=\frac{h^\dagger h + h^T h^*}{32 \pi E},
\end{equation}
with $h=(h_1\, h_2)$ the matrix of Yukawa couplings. Note that decay into both, leptons and antileptons (with total energy $E$), have been included in this expression. As explained e.g. in~\cite{asaka05},  the production and decay terms should be related in order for the right-hand side of eq.~\eqref{ecua_:rho} to vanish in equilibrium, namely
\begin{equation}
\label{ecua_:ratepyd}
\gamma^p= \rho^{\rm eq} \, \Gamma^d \, ,
\end{equation}
with $\rho^{\rm eq}=n^{\rm eq}(t)\, I$ and $I$ the identity matrix, so that the kinetic equation for $\rho$ becomes
\begin{equation}
\label{ecua_:rho2}
\frac{\dif \rho}{\dif t} = -i \conmut{H}{\rho} -\frac{1}{2} \anticonmut{\Gamma^d}{\rho-\rho^{\rm eq}}.
\end{equation}  
Then the equation for the lepton density asymmetry, neglecting washouts, can be obtained taking into account that for each production (destruction) of a neutrino state two leptons or antileptons are destroyed (produced):
\begin{equation*}
\frac{\dif n_L}{\dif t} = -2 \frac{n^{\rm eq}}{32 \pi E} \traza{h^T h^* - h^\dagger h} + 2 \frac{1}{32 \pi E} \traza{[h^T h^* - h^\dagger h] \rho}\, ,
\end{equation*}
where $\traza{.}$ denotes the trace of the corresponding matrix.
Given that $(h^\dagger h)^T=h^T h^*$, the first term, which comes from the destruction of leptons and antileptons, is actually zero, therefore only the production of lepton asymmetry from the decay of the neutrinos contributes to this equation:
\begin{equation}
\label{ecua_:sdma}
\frac{\dif n_L}{\dif t} = 2 \frac{1}{32 \pi E} \traza{[h^T h^* - h^\dagger h] \, \rho}\, \equiv S_{\rm DMA}(t) .
\end{equation}
%%%%%%%%%%%%%%%%%%%%%%%%%%%%%%%%%%%%%%%%%%%%%%%%%%%%%%%%%%%%%%%%%%%%%%%%%%%%%%%
%%%%%%%%%%%%%%%%%%%%%%%%%%%%%%%%%%%%%%
\section{Comparison and discussion}
\label{section_comparison}

To perform an analytical comparison between the EWPA and the DMA, we start by noticing that from
\begin{equation*}
i {\bf G}^{-1}(p^2) = p^2 {\bf 1} - {\bf M^2}(p^2) \simeq i \, {\bf Z}^{-1} \, ({\bf \Delta}^{{\rm BW}})^{-1} \, {\bf Z}^{-1\,T} 
 \end{equation*}
 and ${\bf Z}^{-1} = {\bf Z}^T + \order{h^4}$, one can derive that
 \begin{equation*}
{\bf M^2}(p^2) = {\bf Z}^{T} \,
\begin{pmatrix}
\cpm_{a}^2 & 0 \\
0 & \cpm_{b}^2 
\end{pmatrix} \,
{\bf Z} + \order{h^4, (p^2-\cpm_{a,b}^2)^2} \, .
\end{equation*}
Then, expanding ${\bf M^2}(p^2)$ around $M_{1,2}^2$ we get
\begin{equation*}
{\bf Z}^{T} \,
\begin{pmatrix}
\cpm_{a}^2 & 0 \\
0 & \cpm_{b}^2 
\end{pmatrix} \,
{\bf Z} = 
\begin{pmatrix}
M_{1}^2 & 0 \\
0 & M_{2}^2 
\end{pmatrix}
- \frac{i}{16 \pi}
\begin{pmatrix}
\abss{h_1} & \mire{h_1^* h_2} \\
\mire{h_1^* h_2} & \abss{h_2} 
\end{pmatrix} 
+ \order{h^4, h^2 \epsilon, \epsilon^2} \, .
\end{equation*}
This is the square of the relation we are interested in, namely (using that ${\bf Z}\,{\bf Z}^{T}= I + \order{h^4}$) 
\begin{equation}
\label{ecua_:efhamcomp}
{\bf Z}^{T} \,
\begin{pmatrix}
\cpm_{a} & 0 \\
0 & \cpm_{b}
\end{pmatrix} \,
{\bf Z} = 
H_{\rm ef}
+ \order{h^4, h^2 \epsilon, \epsilon^2} \, ,
\end{equation}
where we have introduced the effective Hamiltonian, 
\begin{equation}
H_{\rm ef} = 
\begin{pmatrix}
M_1 & 0 \\
0 & M_2
\end{pmatrix}
- \frac{i}{32 \pi M}
\begin{pmatrix}
\abss{h_1} & \mire{h_1^* h_2} \\
\mire{h_1^* h_2} & \abss{h_2} 
\end{pmatrix} \, .
\end{equation}

This relation can be used to obtain an approximate expression for the probabilities of  lepton number violating process given in eq.~\eqref{ecua_:prob} that are used in the EWPA  (again taking into account that ${\bf Z}\,{\bf Z}^{T}= I + \order{h^4}$):
\begin{equation}
\label{ecua_:compat}
\begin{split}
\abss{A}(t) &=\frac{1}{(32 \pi \emed)^2} \, \abss{\sum_{j,k} h_j^* h_k^* \left[ {\bf Z}^{T}  
\begin{pmatrix}
e^{-i \left(\pra - i \frac{\pia}{2}\right) \frac{t}{\gamma}} & 0 \\
0 & e^{-i \left(\prb - i \frac{\pib}{2} \right) \frac{t}{\gamma}}
\end{pmatrix}
{\bf Z}\right]_{jk}} \\
&= \frac{1}{(32 \pi \emed)^2} \, \abss{\sum_{j,k} h_j^* h_k^* \left[ \exp\left(-i \, {\bf Z}^{T}
\begin{pmatrix}
\cpm_{a} & 0 \\
0 & \cpm_{b}
\end{pmatrix}
{\bf Z} \, \frac{t}{\gamma}\right)\right]_{jk}} + \order{h^8} \\
&= \frac{1}{(32 \pi \emed)^2} \, \abss{\sum_{j,k} h_j^* h_k^* \left[ e^{-i \, H_{\rm ef} \frac{t}{\gamma}}\right]_{jk}} + \order{h^8, h^6 \epsilon, h^4 \epsilon^2} \\ &
= \frac{1}{(32 \pi \emed)^2} \, \abss{h^* \, e^{-i \, H_{\rm ef} \frac{t}{\gamma}} \, h^{* T}} + \order{h^8, h^6 \epsilon, h^4 \epsilon^2} \, .
\end{split}
\end{equation}
In the same way, for the CP-conjugate process, 
\begin{equation}
\label{ecua_:compbarat}
\abss{\bar A}(t) =  \frac{1}{(32 \pi \emed)^2} \, \abss{h \, e^{-i \, H_{\rm ef} \frac{t}{\gamma}} \, h^T} + \order{h^8, h^6 \epsilon, h^4 \epsilon^2} \, .
\end{equation}
On the other hand, the formal solution to the eq.~\eqref{ecua_:rho2} of the DMA can be written as
\begin{equation}
\rho(t) = \int_0^t \frac{n^{\rm eq}(t')}{32 \pi \emed} \; e^{-i \, H_{\rm ef} \frac{t-t'}{\gamma}} \; \rhop \; e^{i \, H^\dagger_{\rm ef}  \frac{t-t'}{\gamma}} \; \dif t' \, ,
\end{equation}
with $\rhop \equiv h^\dagger h + h^T h^*$. Replacing this expression in the source term of the DMA (i.e. in eq.~\eqref{ecua_:sdma}), some terms cancel and the remaining ones read
\begin{equation}
\label{ecua_:sdmab}
\begin{split}
S_{\rm DMA} & = 2 \int_0^t \frac{n^{\rm eq}(t')}{(32 \pi \emed)^2} \traza{h^T h^*  \, e^{-i \, H_{\rm ef} \frac{t-t'}{\gamma}} \, h^\dagger h \, e^{i \, H^\dagger_{\rm ef}  \frac{t-t'}{\gamma}} - h^\dagger h  \, e^{-i \, H_{\rm ef} \frac{t-t'}{\gamma}} \, h^T h^* \, e^{i \, H^\dagger_{\rm ef}  \frac{t-t'}{\gamma}}} \dif t'\\
& = 2 \int_0^t \frac{n^{\rm eq}(t')}{(32 \pi \emed)^2} \left( \abss{h^* \, e^{-i \, H_{\rm ef} \frac{t-t'}{\gamma}} \, h^{* T}} - \abss{h \, e^{-i \, H_{\rm ef} \frac{t-t'}{\gamma}} \, h^T} \right) \dif t'\, ,
\end{split}
\end{equation}
where we have used the cyclic property of the trace. Therefore, comparing this last equation with eqs.~\eqref{ecua_:sewpa}, \eqref{ecua_:compat} and \eqref{ecua_:compbarat}, we conclude that
\begin{equation}
\label{ecua_:comp}
S_{\rm DMA} = S_{\rm EWPA} + \order{h^8, h^6 \epsilon, h^4 \epsilon^2} \, ,
\end{equation}
i.e. $S_{\rm DMA}$ and $S_{\rm EWPA}$ are equal up to higher order terms in $h^2$ and/or $\epsilon$. 

\begin{figure}[!t]
\centerline{\protect\hbox{
\epsfig{file=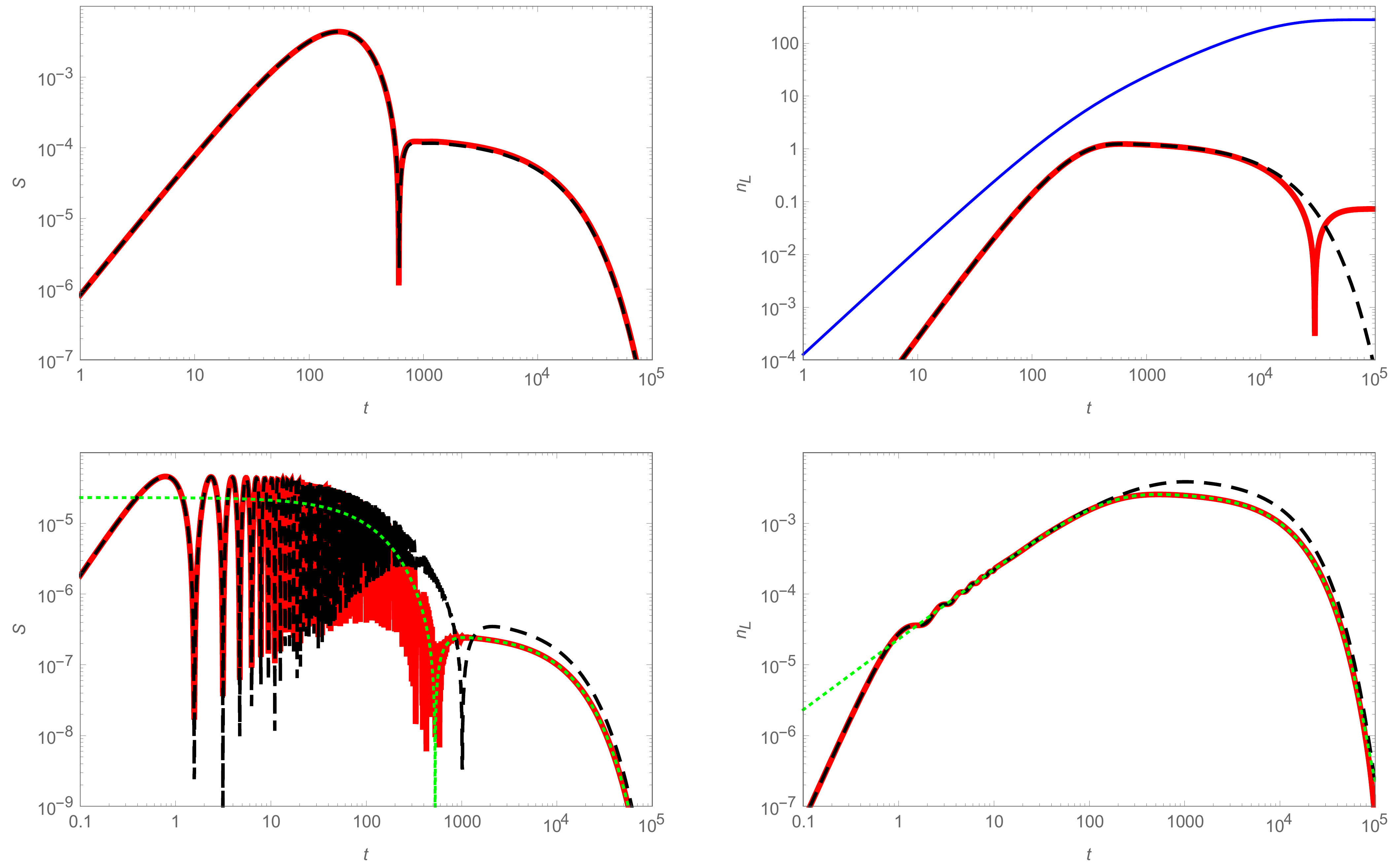,width=0.95\textwidth,angle=0}}} 
\caption[]{Absolute value of the source term (left plots) and lepton asymmetry (right plots), as a function of time normalized to $\gamma/M_1$. The solid red lines correspond to  the EWPA approach, the dashed black lines to the DMA, the blue line in the top right plot to the mixing contribution of the EWPA, and the dotted green lines in the bottom plots to the standard classical approach valid in the hierarchical limit.
In the top plots we have chosen $\Gamma_1/M_1=1/100$, $\Gamma_2/M_1=1/120$ and $\Delta M = \Gamma_1$, 
while for the bottom plots, $\Gamma_1/M_1=1/100$, $\Gamma_2/M_1=1/120$ and $\Delta M/M_1=4$.
 In all cases we have taken $h_1=\abs{h_1}$ and $h_2=\abs{h_2} e^{i \phi}$, with the phase $\phi=\pi/4$, and $n^{\rm eq}(t)=e^{-M_1 \, t/(10000 \, \gamma)}$.  The lepton asymmetry has been obtained integrating only the source term (i.e. ignoring washouts) and the scale on the vertical axis of the plots is not relevant (note that after a change of variables in the integration over time, the factor $M_1/\gamma$ becomes part of the normalization chosen for the lepton asymmetry).} 
\label{figura_:1}
\end{figure}

To corroborate and exemplify this result, we plot in figure~\ref{figura_:1} the evolution of $S_{\rm DMA}$ and $S_{\rm EWPA}$ (left plots), as well as the corresponding lepton asymmetries (right plots), for two different values of the mass splitting $\Delta M \equiv M_2-M_1$. The lepton asymmetries have been obtained by integrating the corresponding source term over time (ignoring washouts) and in both cases we have taken $n^{\rm eq}(t)=e^{-M_1 \, t/(10000 \, \gamma)}$, in order that the time scale of the evolution of $n^{\rm eq}$ be much larger than the lifetimes of the neutrinos. 

The top plots illustrate the highly degenerate case by taking $\Delta M = \Gamma_1$. There are two main things to notice. First, the left plot shows that $S_{\rm DMA}$ and $S_{\rm EWPA}$ are--almost--equal at all times. Second, from the right plot it can be seen that there actually small differences between both sources, coming from the $\order{h^8, h^6 \epsilon, h^4 \epsilon^2}$ terms in eq.~\eqref{ecua_:comp}, 
which accumulate over time and result in noticeable differences between the lepton asymmetries at late times. In particular, while the final lepton asymmetry in the DMA is null, it is not so in the EWPA. This is interesting because unitarity and CPT invariance imply that when washouts are negligible and the initial abundances of neutrinos and lepton asymmetry are null, the final asymmetry must also be null. Therefore the discrepancy between the final lepton asymmetries reflect the non-trivial differences in the--crucial--implementation of unitarity constraints, as explained next.

On one hand, in the kinetic equations of the DMA, as derived above or e.g. in~\cite{Sigl93, asaka05}, the right-hand side of eq.~\eqref{ecua_:rho} is made zero in equilibrium by relating the production and decay rates as in eq.~\eqref{ecua_:ratepyd}. Moreover, the equation for the lepton asymmetry is inferred from some particle number conservation condition. 
It is then easy to show analytically that the final lepton asymmetry in the DMA is {\it exactly} null. Namely, the final lepton asymmetry in the DMA, $n_L^f$, can be obtained integrating the source $S_{\rm DMA}$ given in eq.~\eqref{ecua_:sdma}, 
\begin{equation*}
\begin{split}
n_L^f &=\frac{2}{32 \pi \emed} \int_0^\infty \traza{[h^T h^* - h^\dagger h] \, (\rho-\rho^{\rm eq})} \dif t \\ &=  \frac{2}{32 \pi \emed} \traza{[h^T h^* - h^\dagger h] \,  \int_0^\infty (\rho-\rho^{\rm eq}) \, \dif t} \, ,
\end{split}
\end{equation*} 
where we have taken $n_L(t=0)=0$ and for convenience the null contribution from the terms proportional to $\rho^{\rm eq}$ has been included. Now, from eq.~\eqref{ecua_:rho2} and using that $\conmut{H}{\rho} = \conmut{H}{\rho-\rho^{\rm eq}}$, it is clear that $\int_0^\infty (\rho-\rho^{\rm eq}) \,\dif t = 0$ when the initial and final densities of neutrinos are null, and therefore that $n_L^f$ is zero in the DMA.

This procedure to obtain the kinetic equations in the DMA has sometimes been justified by detailed balance conditions. However, in a time non-invariant theory, as is the case when CPT is conserved but CP is violated, detailed balance may not hold and the canonical equilibrium distributions result from a cyclic balance involving several processes, which is precisely ensured by unitarity (see e.g.~\cite{Dolgov1998, Bernreuther:2002uj} and note that this is at the root of the RIS subtraction procedure to be discussed below). Moreover, this procedure does not clarify whether the rate in eq.~\eqref{ecua_:rated} should be calculated at tree level (which leads to an agreement with the EWPA) or involve effective couplings to account for CP violation in production and decay (due to the presence of oscillatory and secular terms, as well as enhancements inversely proportional to the mass splitting, higher order corrections to the rates could be non-negligible)\footnote{Note that we are not including one-loop vertex corrections, which are negligible in resonant and ARS leptogenesis, and moreover they appear independently in unitarity constraints, see e.g.~\cite{roulet97}.}. 

On the other hand, in the EWPA there is no counting of neutrino abundances, avoiding unitarity problems that may arise when using perturbation theory with unstable particles in the initial or final states. Instead, the source term is built from the probabilities in the eqs.~\eqref{ecua_:prob}, which only involve the stable leptons and antileptons as asymptotic states, and have been obtained after a perturbative expansion of the propagator. Then unitarity constraints are satisfied as a matter of course up to higher order terms in the expansion. Notably, as explained in~\cite{Racker20, Racker21}, there is a non-trivial cancellation between three different terms  contributing to the final lepton asymmetry. These terms (dubbed mixing, oscillation, and interference) arise when calculating the CP asymmetry $\abss{A} - \abss{\bar A}$ from the eqs.~\eqref{ecua_:prob} and will be discussed below in greater detail. To illustrate the cancellation we have plotted one of these contributions in figure~\ref{figura_:1} (the mixing term represented by the blue curve in the top right panel), that should be compared with the net lepton asymmetry (red curve), whose final value is almost null--but not exactly null due to higher order corrections.

In the bottom plots we have taken a very large mass splittings, $\Delta M/M_1=4$, to illustrate the hierarchical limit. As can be seen, there are differences of around 50\% in the lepton asymmetries obtained from the DMA and EWPA. We have verified numerically that the main difference between the left-hand side of eq.~\eqref{ecua_:efhamcomp} and $H_{\rm ef}$ lies in the imaginary part of the diagonal elements. 
Overall, however, the conclusion is that differences are not very significant and only appear at very large mass splittings for which, actually, decoherence effects not included in either approach may become dominant and destroy oscillations (see e.g.~\cite{Beuthe01}). That being said, it is interesting to note that the lepton asymmetry from the EWPA exactly matches the one obtained with the standard classical Boltzmann equations (green curve in figure~\ref{figura_:1} ), 
except at early times (of the order of the oscillation period),
and late times for the same reason as explained above, namely that the classical Boltzmann equations exactly satisfy the unitarity condition of null final asymmetry.
As is well known, in order for the latter to happen it is key to include the rates of some processes subtracting RIS contributions.
Therefore it is reasonable to ask whether this type of RIS-subtraction procedure should also be performed somehow in the DMA in order to be valid in the hierarchical limit (cf.~\cite{Dev:2014laa}). Indeed, as we have shown, such procedure is not required since both, the DMA and the EWPA, which are equivalent up to higher order terms (eq.~\eqref{ecua_:comp}), basically yield the same hierarchical limit as the classical Boltzmann equations (disregarding the minor differences mentioned above). 

In~\cite{Racker20, Racker21} we identified two different types of CP even phases in eqs.~\eqref{ecua_:prob}, one independent of time in $\theta'$, and an oscillating one in the exponentials  $e^{-i  {\tilde M}_j  t/ \gamma}$. Consequently the source term could be written as a sum of three contributions, one from mixing (involving only $\theta'$), one from oscillations (involving only $e^{-i  {\tilde M}_j  t/ \gamma}$), and interference terms (involving both $\theta'$ and $e^{-i  {\tilde M}_j  t/ \gamma}$). This result went along the lines of previous findings~\cite{Dev:2014laa, Dev:2015wpa, Dev:2014wsa, Kartavtsev15}, although differences remained regarding the relative sign between the mixing and oscillation contributions and the existence or not of an interference term. However, 
in the DMA no such separation into CP violation from mixing and oscillations appears (see e.g.~\cite{Klaric:2020lov, Klaric:2021cpi}). Now this can be understood from eqs.~\eqref{ecua_:compat} and~\eqref{ecua_:sdmab}: if the effective Hamiltonian $H_{\rm ef}$ present in the source term of the DMA (eq.~\eqref{ecua_:sdmab}) is diagonalized by an orthogonal matrix, the time exponential involving $H_{\rm ef}$ can be diagonalized by the same matrix, leading to an expression like the first line of eq.~\eqref{ecua_:compat}, where it is possible to identify a time independent CP even phase in the orthogonal matrix and oscillating CP even phases associated to the eigenvalues of $H_{\rm ef}$. Indeed, in the hierarchical limit this decomposition into time independent and oscillating CP even phases becomes quite meaningful, since the oscillation contribution can be identified with the contribution from RIS subtracted rates to the source of the classical Boltzmann equations, as shown in~\cite{Racker20}, while the mixing contribution matches the contribution from decays to the classical source (see also, e.g., \cite{covi96II, Klaric:2021cpi} for related limits).

Moreover, the EWPA is valid for relativistic as well as non-relativistic neutrinos 
and does not involve any counting of neutrino states, therefore it can clearly be applied whether the lepton asymmetry is mostly generated during production or decay. Therefore we think the equivalence we have demonstrated between this approach and the DMA, provides further support to the validity of the DMA for describing jointly  freeze-in an freeze-out leptogenesis, although issues related to the helicity degree of freedom cannot be captured in the scalar toy model (see~\cite{Eijima:2017anv, Ghiglieri:2017gjz, Hambye:2017elz, Ghiglieri:2017csp, Eijima:2018qke, Abada:2018oly, Ghiglieri:2018wbs, Klaric:2020lov, Klaric:2021cpi, Jukkala:2021cys, Drewes:2021nqr}).

Finally, note that in the EWPA, as in other approaches, it is quite involved to obtain  consistent results in the degenerate limit, and even more in the double degenerate limit of equal masses and couplings. Therefore the agreement between the EWPA and the DMA, which does not suffer from this problem, supports the treatment of the degenerate limit in~\cite{Racker21}.

%%%%%%%%%%%%%%%%%%%%%%%%%%%%%%%%%%%%%%
%%%%%%%%%%%%%%%%%%%%%%%%%%%%%%%%%%%%%%%
%%%%%%%%%%%%%%%%%%%%%%%%%%%%%%%%%%%%%%

\section{Conclusions}
\label{section_conclusions}

We have shown analytically that the EWPA and the DMA are equivalent up to higher order corrections in the couplings and mass splitting (see eq.~\eqref{ecua_:comp}). The fact that there are differences coming from higher order terms, actually makes the comparison between both approaches more meaningful, and in particular they reflect the different implementation of unitarity requirements, as discussed in section~\ref{section_comparison}. 
For large mass splittings the lepton asymmetry obtained with the EWPA matches the one from the standard classical Boltzmann equations (except at very early and late times, as explained in section~\ref{section_comparison}), while differences of some tens of percent arise with the DMA. Overall, however, we noted that these differences are not very significant and only appear at very large mass splittings ($\Delta M/M_1=4$ in figure~\ref{figura_:1}), when anyway decoherence effects not included in any of the approaches could become dominant. 

The connection between both approaches established in this work, helps to understand several issues that have been discussed quite extensively in the literature. 
In particular, we have shown that the production and decay rates in the DMA should be calculated at tree level (i.e. without involving effective couplings to account for CP violation in production and decay), in order to get the agreement given by eq.~\eqref{ecua_:comp} with the EWPA, which does not bring this type of doubts. 
Regarding the meaningfulness or existence of different types of contributions to the generation of lepton asymmetry associated to CP violation in mixing, oscillations and interference, we have shown how to get this decomposition from a solution to the kinetic equations of the DMA. In turn, given that in the hierarchical limit the oscillation contribution can be identified with the contribution from RIS subtracted rates to the source of the classical Boltzmann equations (see~\cite{Racker20}), this provides further insight into how the hierarchical limit can be obtained from the DMA without including RIS subtracted rates. Moreover, the validity of the EWPA in the degenerate limit and the double degenerate limit of equal masses and couplings is confirmed by its concordance with the DMA. Finally, the agreement between both approaches also supports their use for a joint analysis of freeze-in and freeze-out leptogenesis, although issues related to the helicity degree of freedom cannot be captured in the scalar toy model we have employed and might be interesting to address in future works. Another possibility worth considering for future research along the lines of this work is the tri-resonant heavy neutrino system~\cite{daSilva:2022mrx}, consisting of three nearly degenerate neutrinos.

%%%%%%%%%%%%%%%%%%%%%%%%%%%%%%%%%%%%%%%%%%%%%%%%%%%%%%%%%%%%%%%%%%%%%%%%%%%
%%%%%%%%%%%%%%%%%%%%%%%%%%%%%%%%%%%%%%%%%%%%%%%%%%%
\bibliographystyle{JHEP}
\bibliography{referencias_leptogenesis3}

\end{document}